\documentclass[sigconf,natbib=true]{acmart}

\copyrightyear{2023}
\acmYear{2023}
\setcopyright{acmcopyright}\acmConference[WSDM '23]{Proceedings of the Sixteenth ACM International Conference on Web Search and Data Mining}{February 27-March 3, 2023}{Singapore, Singapore}
\acmBooktitle{Proceedings of the Sixteenth ACM International Conference on Web Search and Data Mining (WSDM '23), February 27-March 3, 2023, Singapore, Singapore}
\acmPrice{15.00}
\acmDOI{10.1145/3539597.3570365}
\acmISBN{978-1-4503-9407-9/23/02}

\usepackage{afterpage}
\usepackage{booktabs}
\usepackage{multirow}
\usepackage{algorithm}
\usepackage{algorithmic}
\usepackage{enumitem}
\usepackage{caption}
\usepackage{subcaption}
\usepackage{collab}
\collabAuthor{ww}{orange}{Weiwen}
\collabAuthor{rui}{violet}{ZhangRui}

\usepackage{hyperref,tablefootnote,footnotehyper}
\begin{document}

\title{An F-shape Click Model for Information Retrieval on Multi-block Mobile Pages}

\author{Lingyue Fu}
\authornote{Both authors contributed equally to this research.}
\email{fulingyue@sjtu.edu.cn}
\affiliation{
  \institution{Shanghai Jiao Tong University}
  \city{Shanghai}
  \country{China}
}

\author{Jianghao Lin}
\authornotemark[1]
\email{chiangel@sjtu.edu.cn}
\affiliation{
  \institution{Shanghai Jiao Tong University}
  \city{Shanghai}
  \country{China}
}

\author{Weiwen Liu}
\email{liuweiwen8@huawei.com}
\affiliation{
  \institution{Huawei Noah's Ark Lab}
  \city{Shenzhen}
  \country{China}
}

\author{Ruiming Tang}
\email{tangruiming@huawei.com}
\affiliation{
  \institution{Huawei Noah's Ark Lab}
  \city{Shenzhen}
  \country{China}
}

\author{Weinan Zhang}
\authornote{The corresponding authors.}
\email{wnzhang@sjtu.edu.cn}
\affiliation{
  \institution{Shanghai Jiao Tong University}
  \city{Shanghai}
  \country{China}
}

\author{Rui Zhang}
\email{rayteam@yeah.net}
\affiliation{
  \institution{ruizhang.info}
  \city{Shenzhen}
  \country{China}
}

\author{Yong Yu}
\authornotemark[2]
\email{yyu@sjtu.edu.cn}
\affiliation{
  \institution{Shanghai Jiao Tong University}
  \city{Shanghai}
  \country{China}
}



\renewcommand{\shortauthors}{Lingyue Fu and Jianghao Lin et al.}

\begin{abstract}
To provide click simulation or relevance estimation based on users' implicit interaction feedback, click models have been much studied during recent years.
Most click models focus on user behaviors towards a single list. 
However, with the development of user interface (UI) design, the layout of displayed items on a result page tends to be multi-block (i.e., multi-list) style instead of a single list, which requires different assumptions to model user behaviors more accurately. 
There exist click models for multi-block pages in desktop contexts, but they cannot be directly applied to mobile scenarios due to different interaction manners, result types and especially multi-block presentation styles. 
In particular, multi-block mobile pages can normally be decomposed into interleavings of basic \emph{vertical blocks} and \emph{horizontal blocks}, thus resulting in typically \emph{F-shape} forms.
To mitigate gaps between desktop and mobile contexts for multi-block pages, we conduct a user eye-tracking study, and identify users' sequential browsing, block skip and comparison patterns on F-shape pages.
These findings lead to the design of a novel F-shape Click Model (FSCM), which serves as a general solution to multi-block mobile pages.
\emph{Firstly}, we construct a Directed Acyclic Graph (DAG) for each page, where each item is regarded as a vertex and each edge indicates the user's possible examination flow. 
\emph{Secondly}, we propose DAG-structured GRUs and a comparison module to model users' sequential (sequential browsing, block skip) and non-sequential (comparison) behaviors respectively.
\emph{Finally}, we combine GRU states and comparison patterns to perform user click predictions.
Experiments on a large-scale real-world dataset validate the effectiveness of FSCM on user behavior predictions
compared with baseline models.
\end{abstract}



\maketitle

\section{Introduction}
\label{sec:intro}


Modeling user behaviors is key to improving the performance of information retrieval systems. The ability to accurately capture user behaviors allows a retrieval system to better fulfill users' information needs~\cite{lin2021graph}. To this end, many \emph{click models} are proposed to model users’ click behaviors. They serve as click simulators in cases where no real users are available or when we prefer not to experiment with real users to avoid hurting user experiences~\cite{borisov2016neural}. Besides, click models are also used to provide relevance estimations~\cite{chen2020context,wang2013content}.

\begin{figure}[t]
    \centering
    \includegraphics[width=0.47\textwidth]{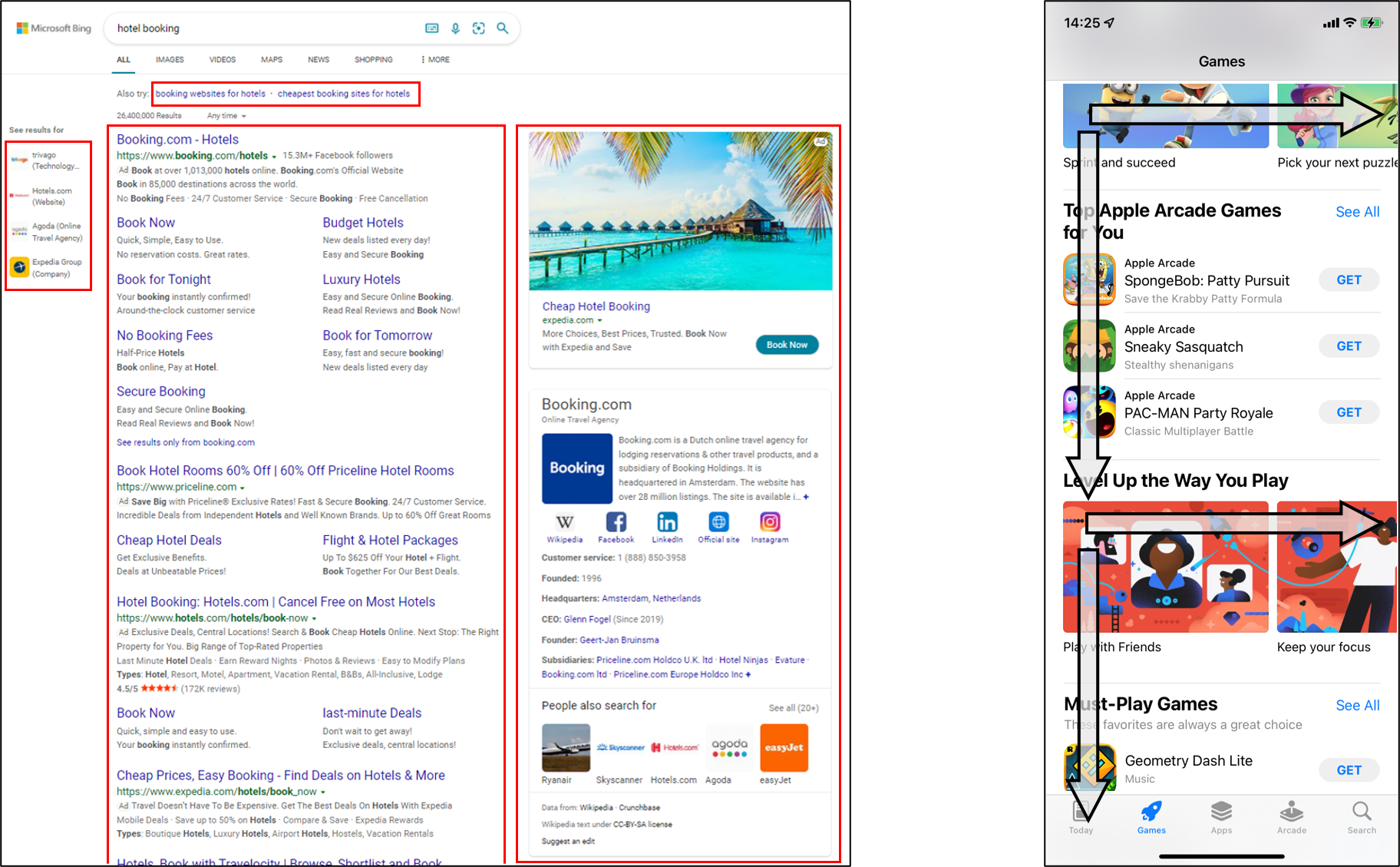}
    \caption{Multi-block pages for desktop (left) and mobile (right) contexts. Left: a result page from Bing Search, where each block is indicated by a red box. Right: a result page from App Store, which can be decomposed into basic vertical and horizontal blocks, thus resulting in an F-shape form.}
    \label{fig:illustration}
    \vspace{-15pt}
\end{figure}

The performance of a click model heavily depends on appropriate assumptions on users' behavior patterns to tackle different biases in click logs (e.g., position bias~\cite{craswell2008experimental}, appearance bias~\cite{wang2013incorporating}). 
Earlier click models focus on user behaviors towards a single list presented on a search engine result page (SERP) in the desktop context. 
Most of them follow the examination hypothesis~\cite{richardson2007predicting}, where a user clicks an item if and only if she examines the item and is attracted by the item. \citet{craswell2008experimental} propose Cascade Model (CM) which assumes that users scan each item in the list from top to bottom until the first click. Expanding upon Cascade Model, Dependent Click Model (DCM)~\cite{guo2009efficient} assumes that users have a certain probability to examine the next item after clicking the current item. 
Comparison-based Click Model (CBCM)~\cite{zhang2021constructing} further considers non-sequential browsing behaviors (e.g., revisit, compare). Besides, recent works~\cite{borisov2016neural,chen2020context,dai2021adversarial,lin2021graph} start to introduce neural networks to click models to enable automatic dependency detection and larger model capacities.

Despite the great attempts in user modelings on desktop pages, researchers point out that user behaviors in mobile environments are different from those in desktop contexts, which requires refinement for existing behavioral assumptions~\cite{mao2018constructing,verma2016characterizing}. 
For example, users pay more attention to the top-ranked results and scan fewer results on small screens~\cite{kim2015eye}.
\citet{mao2018constructing} propose Mobile Click Model (MCM) to incorporate click necessity bias and examination satisfaction bias in mobile search. Viewport Time Click Model~\cite{zheng2019constructing} further extends MCM by utilizing viewport time on mobile screens.

However, the above-mentioned clicks models (either for desktop or mobile environments) only consider user behavior patterns towards a single list.
Nowadays, with the development of user interface (UI) design, the layout of displayed items on a result page tends to be multi-block presentation style instead of a single list~\cite{xi2023bird}. 
As shown in Figure~\ref{fig:illustration}, typical multi-block displays for desktop and mobile environments are quite different. While multi-block desktop pages are usually unaligned and contain blocks of various sizes, multi-block mobile pages can be generally decomposed into two basic blocks (i.e., \emph{vertical blocks} and \emph{horizontal scrolling blocks}), thus resulting in typically \emph{F-shape} forms.
The aforementioned click models only model user behaviors exclusively for each single block as a list, which sacrifices useful browsing information embedded in other blocks, leading to inaccurate user modeling. 

There exist click models for multi-block pages in the desktop environment. Joint Relevance Examination Model (JRE)~\cite{srikant2010user} and General Click Model (GCM)~\cite{zhu2010novel} interpret user behaviors in ads blocks. Whole Page Click Model (WPC)~\cite{chen2011whole} adopts the Markov decision process to model the examination and the skip among blocks. Nevertheless, these multi-block click models only consider examination transferring among blocks and ignore the interactive influence across different blocks, leading to poor user modeling performance. More importantly, different multi-block styles shown in Figure~\ref{fig:illustration}, together with different interaction manners and result types, lead to different user behaviors in desktop and mobile contexts (more details can be found in Section \ref{sec:user behavior}). These gaps make it impractical and ineffective to simply adopt desktop-oriented multi-block click models to mobile scenarios. Hence, user behavior patterns on multi-block mobile pages need further investigation.

To explore this, we conduct a lab-based eye-tracking study towards \emph{F-shape} pages. 
We argue that F-shape pages are prevalent and representative for multi-block mobile pages. F-shape pages contain interleavings of two basic mobile elements (i.e., vertical and horizontal blocks) and can be found on many popular Apps like \href{https://www.amazon.com/}{Amazon}, \href{https://www.youtube.com/}{YouTube}, \href{https://medium.com/}{Medium} and \href{https://www.pandora.com/}{Pandora}, as well as App stores provided by \href{https://www.apple.com/}{Apple}, \href{https://consumer.huawei.com/en/phones/}{Huawei}, \href{https://www.samsung.com/us/smartphones/}{Samsung}, etc. 
Based on the eye-tracking data, we identify primary principles for users' \textbf{sequential browsing}, \textbf{block skip} and \textbf{comparison} patterns on F-shape pages.
Motivated by these findings, we propose a novel F-shape Click Model (FSCM). 
\emph{Firstly}, we construct a directed acyclic graph (DAG) for each F-shape page, where each item on the page is regarded as a vertex and each directed edge indicates the user's possible examination flow. 
\emph{Secondly}, we propose DAG-structured gated recurrent units (GRUs) to model users' sequential browsing and block skip behaviors. A comparison module is further designed for users' comparison patterns.
\emph{Finally}, we combine the GRU states and comparison patterns through an output layer to perform user click prediction. 
\emph{Moreover}, FSCM serves as a general solution and can flexibly adapt to multi-block mobile pages that break F-shape forms (e.g., consecutive horizontal blocks). 
The main contributions of this paper are:
\begin{itemize}[leftmargin=10pt]
    \item We carry out an eye-tracking study to investigate user behaviors on F-shape pages, and identify primary principles for users' \emph{sequential browsing}, \emph{block skip} and \emph{comparison} patterns. Raw data and source codes of eye-tracking experiments are open-sourced to encourage future works in related research communities.
    \item We propose a novel F-shape click model (FSCM). We apply DAG-structured GRUs and a comparison module to model users' sequential (sequential browsing, block skip) and non-sequential (comparison) behaviors respectively.
    To the best of our knowledge, this is the first general solution to address user modeling problems on multi-block mobile pages. 
    \item Extensive experiments show that FSCM significantly outperforms baseline click models. This validates that the underlying assumptions of FSCM are closer to practical user behaviors than the assumptions made in competing models.
\end{itemize}
\section{Related Works}

\subsection{User Behaviors on Mobile}
\label{sec:user behavior}

With the rise of smartphones and mobile search, understanding user behaviors on mobile devices becomes increasingly important. Researchers have characterized differences in user behavior patterns between desktop and mobile in various aspects. 

First, the user interface (UI) of mobile environment is very different from that of desktop context. Unlike a desktop computer with a large screen as well as a mouse and a keyboard as input devices, a mobile phone often has a much smaller screen and view users' touch interactions (e.g., swiping, tapping) as input signals, leading to different behavior patterns~\cite{zheng2019constructing}. Regarding different input interactions, \citet{guo2013mining} utilize mobile touch interactions to estimate the relevance of displayed items, and identify characteristics of users' fine-grained interactions on landing pages. Besides, smaller device screens impose more efforts for mobile users to gather the same amount of information. \citet{kim2015eye} conduct an eye-tracking study to find that users put more attention to top-ranked results and exhibited a more linear scanning pattern on mobile screens. \citet{wu2014using} and \citet{ong2017using} find that users use different browsing strategies to adapt to result pages with various Information Scent Levels and Information Scent Patterns.

Second, compared to desktop environments, information needs and contexts of mobile users tend to be more diverse and changeable~\cite{kamvar2009computers,yi2008deciphering}. \citet{song2013exploring} suggest that the information needs and click preference of mobile users vary across many different factors, including the time of the day, locations, and search devices. \citet{harvey2017searching} find that mobile users often search in an “on the go" context, where they might be interrupted or distracted. A user study is further conducted to emphasize the impact of “fragmented attention" on behavior patterns and model performance. 

These studies suggest that users' behaviors on mobile devices are different from those in desktop settings, thus requiring refinements on click models that are originally designed for desktop contexts. In this work, we focus on user behaviors on multi-block mobile pages, which have not been studied yet. By conducting eye-tracking experiments and data analysis, we propose an F-shape click model (FSCM) to characterize user behaviors among blocks, and achieve page-level click predictions and optimizations.

\begin{figure}[t]
    \centering
    \includegraphics[width=0.44\textwidth]{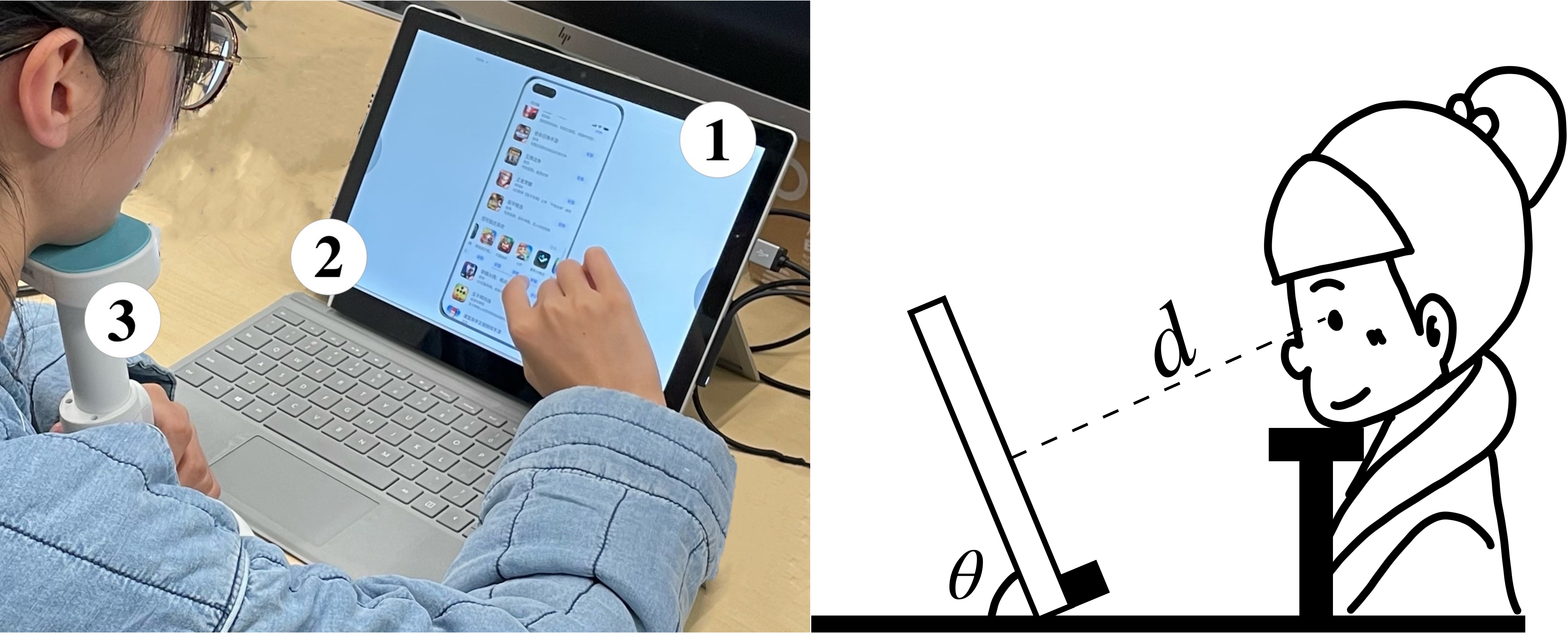}
    \vspace{-5pt}
    \caption{Left: eye-tracking experimental setups, depicting (1) the screen where the pseudo mobile UI is displayed, (2) the eye-tracker and (3) the chin rest to restrain head movements.
    Right: illustrations of how our setting preserves the natural viewing distance and viewing angle in mobile scenarios.}
    \label{fig:eye-tracking setup}
    \vspace{-13pt}
\end{figure}

\subsection{Click Models}

To model and simulate user behaviors, numerous click models have been proposed for various application scenarios (e.g., web search, recommendation)~\cite{chuklin2015click,zhong2010incorporating,xu2012incorporating,chen2020beyond}. Earlier click models are based on the probabilistic graphical model (PGM) framework and mainly focus on user behaviors towards a single list in the desktop or mobile contexts. They usually adopt different assumptions on user behaviors to specify how items and clicks at different positions affect each other. The most common assumption is the examination hypothesis~\cite{richardson2007predicting} where the click probability is decomposed into the examination probability and attractiveness score. 
A simple click model that follows the examination hypothesis is the Position-based Model (PBM)~\cite{craswell2008experimental}, which assumes that the examination probability is only related to the displayed position. The Cascade Model (CM)~\cite{craswell2008experimental} assumes that users scan each document in the list from top to bottom until the first click. Expanding upon the cascade model, User Browsing Model (UBM)~\cite{dupret2008user}, Dynamic Bayesian Network (DBN)~\cite{Chapelle2009DBN}, Dependent Click Model (DCM)~\cite{Guo2009DCM}, and Click Chain Model (CCM)~\cite{Guo2009CCM} have been proposed to make additional assumptions and promote the model performance. Moreover, neural network (NN) based methods~\cite{borisov2016neural,borisov2018click,dai2021adversarial,chen2020context,lin2021graph} are proposed for better expressive power and flexible dependencies. They treat user behaviors as a sequence of hidden state vectors and adopt recurrent neural networks to model users' dynamic interaction patterns.

The aforementioned click models only focus on user behavior patterns towards a single list. However, with the development of user interface (UI) design, the layout of displayed items on a result page tends to be multi-block presentation style instead of a single list, which requires further analysis and modeling. For multi-block desktop pages, the Joint Relevance Examination Model (JRE)~\cite{srikant2010user} and General Click Model (GCM)~\cite{zhu2010novel} consider behavior patterns within ads blocks and enhance predictions on main result blocks.
Whole Page Click Model (WPC)~\cite{chen2011whole} adopts Markov decision process to model examinations and skips among multiple blocks, and achieves better performance on page-level click predictions. 

As discussed above, different multi-block styles shown in Figure 1, together with different interaction manners and result types, lead to different behavior patterns in desktop and mobile contexts. These facts make it impractical and ineffective to simply adopt multi-block click models for desktop contexts to mobile scenarios. Therefore, click models for multi-block mobile pages need further investigation. To the best of our knowledge, we are the first to study user behaviors on multi-block mobile pages. Our proposed F-shape Click Model (FSCM) can better extract interactive effects among blocks and thus achieve the state-of-art performance compared with existing click models.


\section{Eye-Tracking and Data Analysis}
\label{sec:eye tracking}

To study user behaviors towards multi-block mobile pages, we  carried out a laboratory eye-tracking experiment with 20 participants (20 participants are comparable to previous works~\cite{buscher2009you,peitek2018simultaneous}). We choose F-shape pages as the representative multi-block mobile pages as stated in Section~\ref{sec:intro}.

Each participant is asked to examine $20$ F-shape pages collected from a mainstream commercial App store. There are 6 items for each vertical block and 8 items for each horizontal block. Since user behaviors can be affected by different types of information needs, we divide 20 F-shape pages into 10 exploratory tasks and 10 informational tasks. 
In exploratory tasks, participants are asked to explore the whole page without any purpose and click whatever they like.
In informational tasks, participants are first told a keyword (e.g., video players, music apps) and then given a relevant search result page from the App store. They are required to browse the page and click items that they are satisfied with.


The settings of our eye-tracking experiment
are shown in Figure~\ref{fig:eye-tracking setup}. Following previous works~\cite{leiva2020understanding,hofmann2014eye,zhang2021constructing}, we present each page via a pseudo front end on a laptop with a touchscreen to collect the eye-tracking data. It is worth noting that participants can only utilize the touchscreen to interact with the pages (e.g., swiping, tapping) to simulate mobile user interactions. 

\begin{figure*}
     \centering
     \begin{subfigure}[b]{0.2\textwidth}
         \centering
         \includegraphics[width=\textwidth]{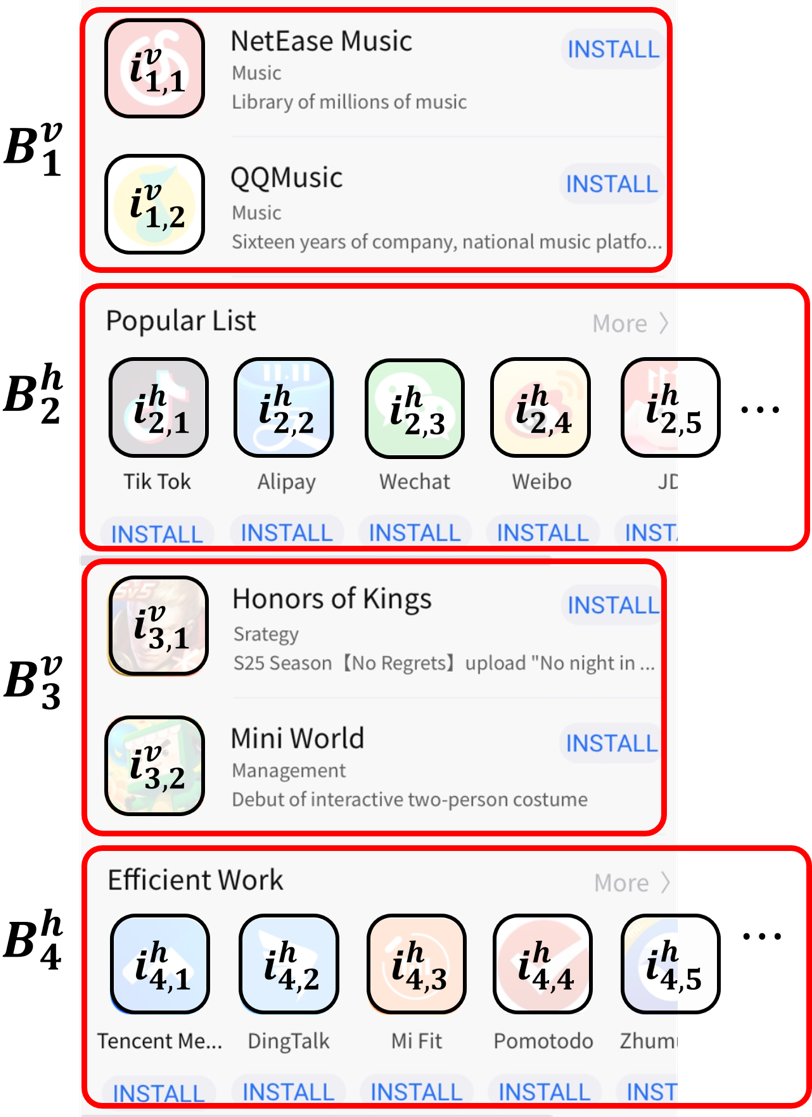}
         \caption{}
         \label{subfig:eye-tracking-notation}
     \end{subfigure}
     \begin{subfigure}[b]{0.262\textwidth}
         \centering
         \includegraphics[width=\textwidth]{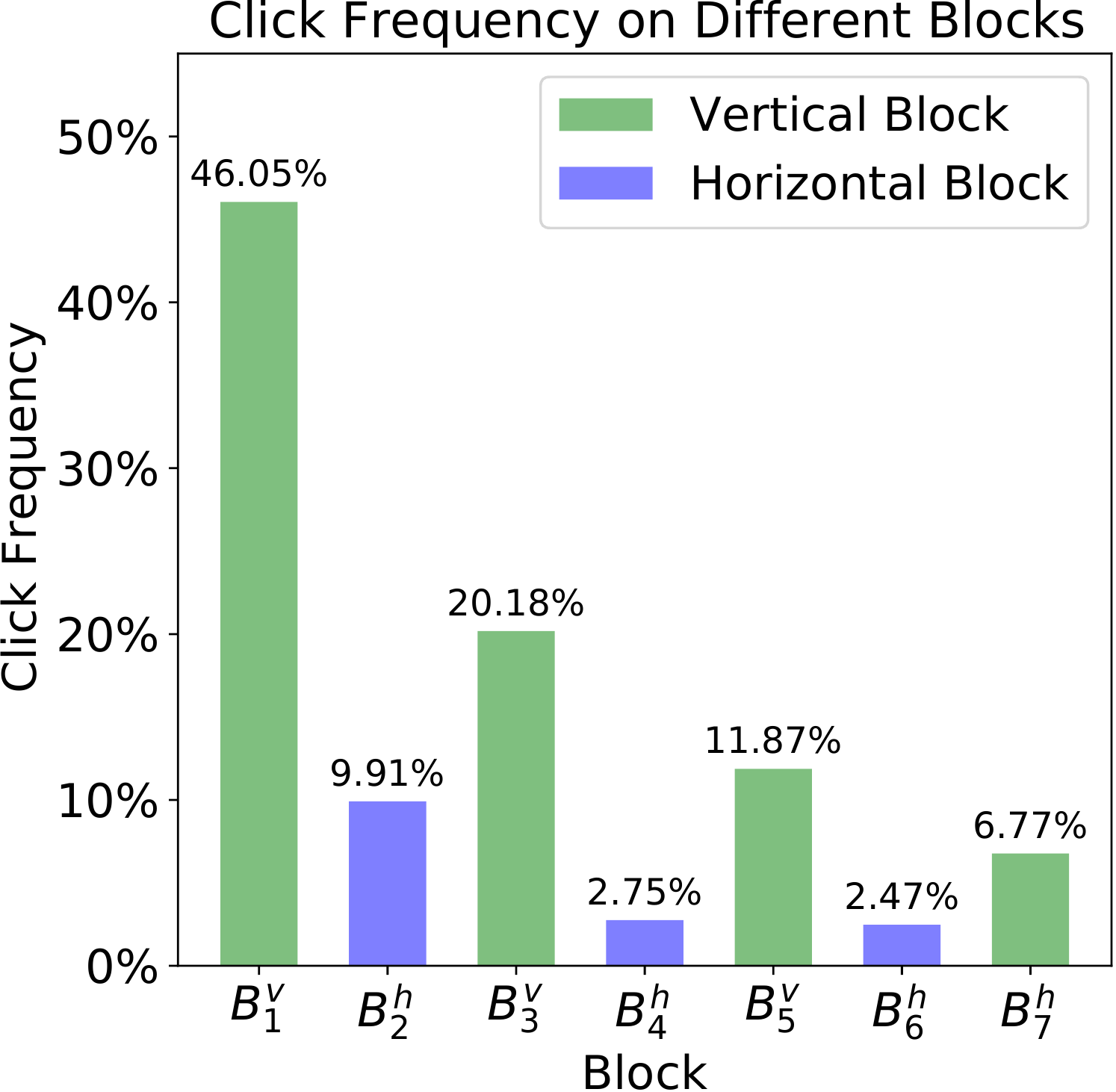}
         \caption{}
         \label{subfig:click frequency on different blocks}
     \end{subfigure}
     \begin{subfigure}[b]{0.262\textwidth}
         \centering
         \includegraphics[width=\textwidth]{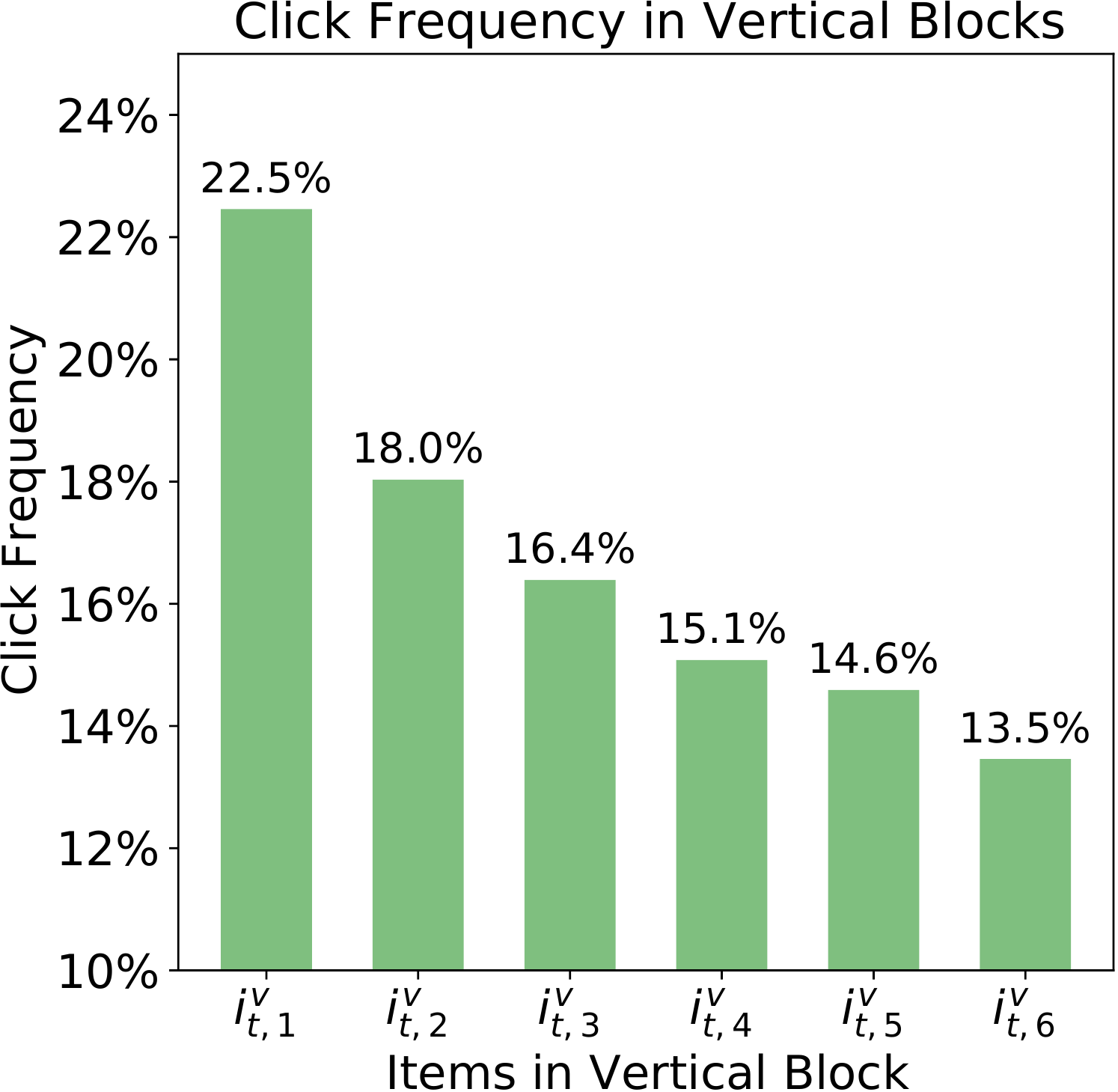}
         \caption{}
         \label{subfig:click frequency in vertical blocks}
     \end{subfigure}
     \begin{subfigure}[b]{0.262\textwidth}
         \centering
         \includegraphics[width=\textwidth]{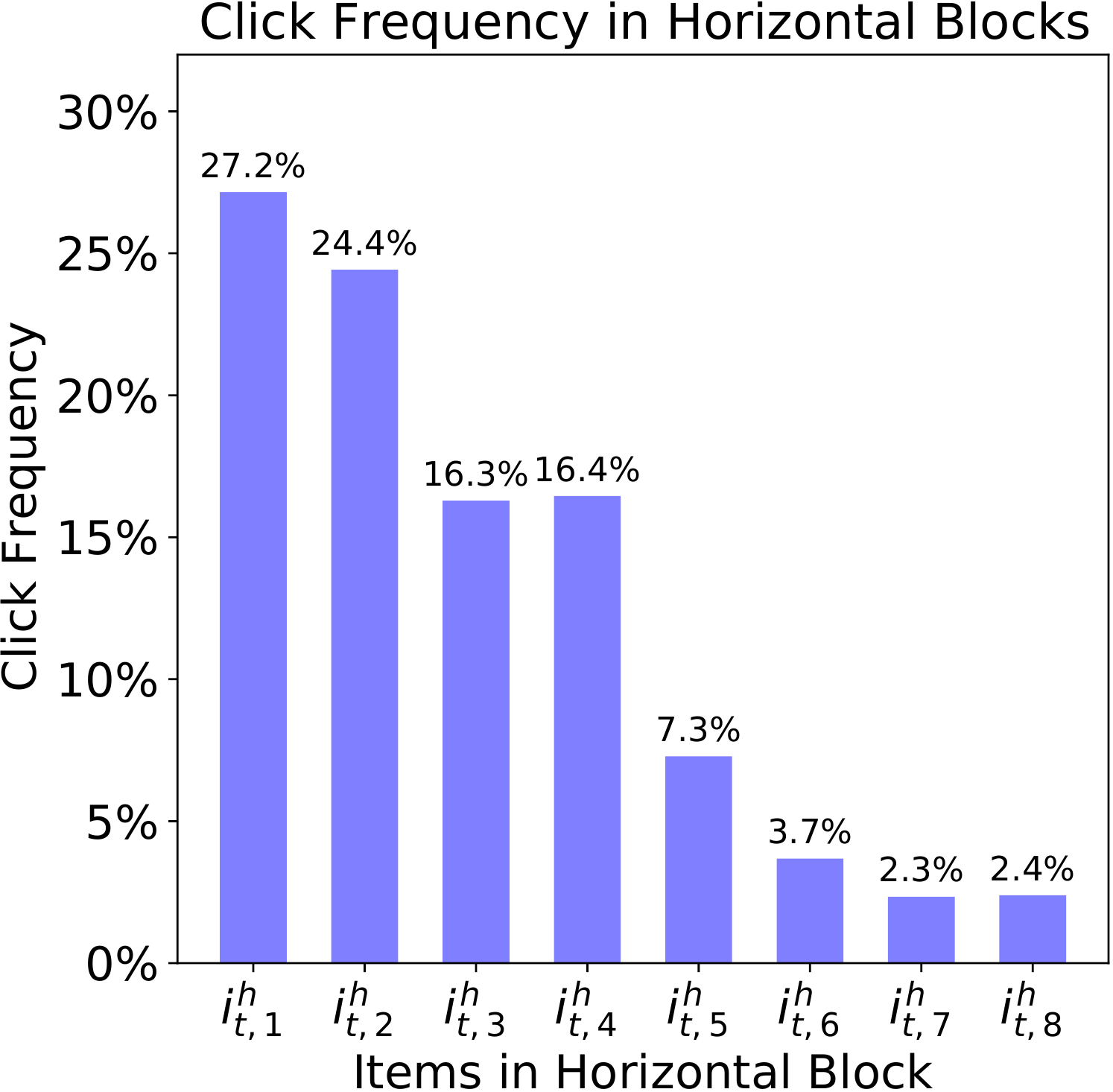}
         \caption{}
         \label{subfig:click frequency in horizontal block}
     \end{subfigure}
     \vspace{-20pt}
     \caption{(a) Decomposition of an F-shape page. We shrink the length of vertical blocks due to page limitations. (b) Click frequency on different blocks. (c) Click frequency on items in vertical blocks. (d) Click frequency on items in horizontal blocks.}
     \label{fig:eyetracking-notation and click frequency}
\end{figure*}

We use Tobii 4C \footnote{\url{https://gaming.tobii.com/products/}} to collect eye-tracking data. We deploy our pseudo front end and Tobii driver on Microsoft Surface Pro 5.
The laptop screen is 12.3 inches wide with resolution of $2736 \times 1824$ in pixels. During experiments, participants sit comfortably in front of the laptop. According to \cite{long2017viewing,leiva2020understanding}, we keep the distance between participants' eyes and the screen within 30-40 cm range.
The laptop is tilted at an angle $\theta$ in the range of 65-70 degree to preserve the viewing angle of mobile scenarios. Before the experiment, we calibrate the Tobii device, and then use a chin rest to  restrain participants' head movement during the experiments. After the experiment, we test the Tobii device again to verify its correctness. In this way, the error caused by the eye tracker itself does not exceed $0.57$ degrees in all directions in our experiments.

We collect a total of $400$ results from 20 participants on 20 pages, of which $346$ are valid, accounting for $86.5\%$. Invalid results are caused by hardware or software faults, and user inattention (i.e., the fraction of user's fixation on the viewport of pseudo mobile UI is less than $90\%$). In those valid results, the average fraction of fixation on mobile UI is 97.7$\%$.
The raw data and source code of eye-tracking experiments are publicly available\footnote{Raw data and source code of eye-tracking experiments: \url{https://bit.ly/3o9mtg2}} and will be open-sourced upon the acceptance of this work.
Equipped with eye-tracking and click data, we aim to answer the following research questions:

\begin{itemize}
    \item[\textbf{RQ1}] Does position bias exist on F-shape pages?
    \item[\textbf{RQ2}] Do users skip some blocks during the sequential browsing?
    \item[\textbf{RQ3}] Do users compare adjacent items when non-sequential examination happens?
\end{itemize}

By investigating these three questions, our goal is to obtain user behavior patterns towards multi-block mobile pages in sequential (\textbf{RQ1} and \textbf{RQ2}) or non-sequential (\textbf{RQ3}) manners. The results can be adopted to model users’ behaviors and improve the performance of click models. To formulate these problems more conveniently, we will first give some necessary definitions and notations before elaborating on detailed analysis.

As shown in Figure~\ref{subfig:eye-tracking-notation}, an F-shape page is decomposed into vertical and horizontal blocks. We denote the blocks from top to bottom as $\mathcal{B}=[B_1^v,B_2^h,B_3^v,B_4^h,\ldots,B_n^v]$, where $v$ and $h$ stand for vertical and horizontal blocks.
Each block $B_t^{b}$ ($b \in \{v,h\}, t=1,2,\ldots,n$)
contains $m_t$ items. We denote items of block $B_t^{b}$  from top to bottom ($b = v$) or from left to right ($b = h$) as $\mathcal{I}_t^{b}=[i_{t,j}^b]_{j=1}^{m_t}$, where $i_{t,j}^b$ is the $j$-th item of the $t$-th block on the page.
For each page, a user's examination sequence is defined as $E=[E_1,E_2,\ldots,E_m]$. Each element $E_k=(t,j)$ indicates the item of the user's $k$-th examination (i.e., $i_{t,j}^b$).
During the examination, users browse each item in the sequence $E$ chronologically,
and an item can be examined multiple times. The user's continuous fixation of the same item is regarded as one examination, which is expressed as: $E_k \neq E_{k-1},2 \le k \le m$.

\begin{table}[tb]
    \centering
    \caption{Block-skip behaviors with different skip lengths.}
    \vspace{-5pt}
	\resizebox{0.4\textwidth}{!}{
    \begin{tabular}{cccccc}
      \toprule
      Skip Length & 2 & 3 & 4 & 5 & 6 \\
        \midrule
      Number & 439&  34&  17&  1&  2\\
      Proportion & 89.05\% & 6.90\% & 3.45\% & 0.20\% & 0.41\% \\
        \bottomrule
    \end{tabular}
    }
    \label{tab:proportion of block-skip behavior}
\end{table}

\subsection{RQ1: Position Bias}

The position bias refers to the fact 
that users are more inclined to interact with the top ranked items in a single list~\cite{guan2007eye,joachims2017accurately,lorigo2008eye}. We aim to explore whether this prior principle is still applicable when extended to multi-block F-shape pages. 

We first focus on \emph{block-level} position bias and calculate the click frequency over different blocks. The results are shown in Figure~\ref{subfig:click frequency on different blocks}. We observe that: (1) For the same type of blocks, users tend to interact with blocks at higher ranks. (2) Users prefer interacting with vertical blocks, even when they are placed at lower ranks than horizontal blocks. The possible reason is the presentation bias~\cite{yue2010beyond}, where vertical blocks contain more descriptions and features, and thus become more attractive to users. These observations reveal that position bias still exists at block level, and can be somehow affected by contents and layouts of the blocks.

Furthermore, we study \emph{intra-block} position bias inside each block and calculate the averaged click frequency on items at different positions in vertical or horizontal blocks. The results are shown in Figure~\ref{subfig:click frequency in vertical blocks} and~\ref{subfig:click frequency in horizontal block}. We find that click frequency gradually decreases as the item position increases inside a block, which means that the position bias still holds inside each individual block. 

In conclusion, there exist block-level and intra-block position bias on F-shape pages. And the block-level position bias can be affected by different types of blocks, where users are more inclined to interact with vertical blocks than horizontal blocks. These conclusions help us identify users' basic examination flows on an F-shape page: users generally browse the F-shape page in top-to-bottom and left-to-right manners. By identifying these two browsing manners, we define the \emph{sequential examination behavior} on F-shape pages as two consecutive items $E_k=(t_1,j_1)$ and $E_{k+1}=(t_2,j_2)$ in the examination sequence $E$ that satisfy $(t_1 < t_2) \lor ((t_1 = t_2) \land (j_1 < j_2))$. 
It is worth noting that we define users' sequential examination behaviors in a general perspective. 
Characteristics related to page contents should be further considered in specific click model designing phases (e.g., presentation bias that causes different click distributions between vertical and horizontal blocks).



\vspace{-5pt}
\subsection{RQ2: Block Skips}

Some blocks may be just skipped and will not be examined by the user during the sequential browsing. This research question aims to study the characteristics of users' block-skip behaviors. We define the \emph{$l$-length block-skip behavior} as two consecutive items $E_k=(t_1,j_1)$ and $E_{k+1}=(t_2,j_2)$ in the examination sequence $E$ that satisfy $l = t_2 - t_1 \ge 2$. We also define $l = t_2 - t_1$ as the \emph{skip length}.

\begin{table}[tb]
    \centering
    \caption{Proportions of different positions as the source or destination for V-V 2-length block-skip behaviors. 
    }
    \vspace{-7pt}
	\resizebox{0.48\textwidth}{!}{
    \begin{tabular}{ccccccc}
      \toprule
      Position & 1 & 2 & 3 & 4 & 5 & 6 \\
        \midrule
      Source Proportion & 0.69\% &  4.17\% &  6.94\% &  6.94\% &  25\% &  56.25\%  \\
      Destination Proportion & 47.2\% & 20.83\% & 9.72\% & 6.25\% & 7.64\% & 8.33\% \\
        \bottomrule
    \end{tabular}
    }
    \vspace{-10pt}
    \label{tab:proportion of src and dst for V-V block skip}
\end{table}

First, we count the number of block-skip behaviors with different skip lengths. Results are summarized in Table~\ref{tab:proportion of block-skip behavior}. There are a total of 493 block skips out of 346 valid pages, which implies that the average number of block skips per page is larger than 1.
Then, we observe that most block-skip behaviors take place with skip length $l=2$. Since our F-shape pages are interleavings of vertical and horizontal blocks, there are two cases for $2$-length block-skip behavior: (a) V-V, from vertical to vertical (b) H-H, from horizontal to horizontal. We find that V-V block skips are dominant (94.5\%) among $2$-length block-skip behaviors, compared with H-H block skips (5.5\%). This reveals that users are more likely to skip horizontal blocks instead of vertical blocks, which is consistent with the previous observation that users are more inclined to interact with vertical blocks.

Next, focusing on V-V block-skip behavior with skip length 2, we aim to study at what position will the user perform a V-V block-skip behavior (source) and which position in the next vertical block will this skip land on (destination).
We calculate the proportion of different positions as the source or destination for a V-V block-skip behavior. The results are shown in Table~\ref{tab:proportion of src and dst for V-V block skip}. We find that V-V block-skip behaviors are likely to take place at the end of a vertical block (56.25\% at position 6) and land on the beginning of the next vertical block (47.2\% at position 1). For instance, as illustrated in Figure~\ref{subfig:eye-tracking-notation}, users are most likely to skip from the last item of $B_1^v$ (i.e., $i_{1,2}^v$) to the first item of $B_{3}^v$ (i.e., $i_{3,1}^v$).

In summary, users do skip some blocks during the sequential browsing, and block skips are mostly V-V $2$-length block-skip behaviors, taking place from the end of one vertical block to the beginning of the next vertical block.







\subsection{RQ3: Comparison Behaviors}

We have studied users' sequential examination behaviors in \textbf{RQ1} and \textbf{RQ2}. Many previous works~\cite{xu2012incorporating,wang2015incorporating,borisov2018click} suggest that users revisit items when browsing a result page, which is referred to as non-sequential behaviors.~\citet{zhang2021constructing} conduct eye-tracking experiments and further point out a crucial pattern among those non-sequential behaviors, named comparison behaviors. They claim that users make click-through decisions by comparing adjacent items, instead of in an isolated manner. The \emph{comparison behavior} can be defined as three consecutive items (triplet) $E_k$, $E_{k+1}$ and $E_{k+2}$ in the examination sequence $E$ that satisfy $E_{k}=E_{k+2}$. That is, the user's attention quickly shifts between $E_k$ and $E_{k+1}$ to make comparisons, possibly resulting in a click feedback.

Following the definition of comparison behavior, we count the number of times over different combinations of the triplet items in users' examination sequences $E$. The results are shown in Table~\ref{tab:comparison behavior}, where $A=(t_a,j_a)$, $B=(t_b,j_b)$, $C=(t_c,j_c)$ satisfy the definition of sequential examination behavior. That is, $(t_a < t_b)\lor((t_a = t_b) \land (j_a < j_b))$, and $(t_b < t_c)\lor((t_b = t_c) \land (j_b < j_c))$ are both true.
We observe that sequential triplet $[A, B, C]$ is dominant, which is in line with the intuition that users generally browse the page in a sequential manner. 
As for non-sequential triplets, the most frequent combinations are $[A,B,A]$ and $[B,A,B]$, which correspond to comparison behaviors (attention shifts between item A and B). This validates that comparison patterns are prevalent among non-sequential behaviors and need further consideration when designing a click model, which is consistent with previous studies~\cite{zhang2021constructing}.
Therefore, we can conclude that users are likely to make comparisons when non-sequential examination happens. 
\begin{table}[tb]
    \centering
    \caption{Count and frequency of combinations over the triplet items in users' examination sequences $E$. 
    }
    \vspace{-5pt}
	\resizebox{0.4\textwidth}{!}{
    \begin{tabular}{cccccc}
      \toprule
        Triplet Type & 1st & 2nd& 3rd & Count & Freq. \\
        \midrule
        Seq. & A& B & C & 5771 & $43.03\%$\\
        Non-seq. & B&A&B &1668	& $12.44\%$\\
        Non-seq. &A&B&A& 1627	& $12.13\%$\\
        Non-seq. &C&B&A& 1320	&$9.84\%$\\
        Non-seq. &B&A&C &858&	$6.40\%$\\
        Non-seq. &A&C&B&846&	$6.31\%$\\
        Non-seq. &B&C&A&705&	$5.26\%$\\
        Non-seq. &C&A&B& 615	& $4.59\%$\\
    \bottomrule
    \end{tabular}
    }
    \vspace{-15pt}
    \label{tab:comparison behavior}
\end{table} 

By answering the three research questions above, we have identified users' \emph{sequential browsing}, \emph{block skip} and \emph{comparison} patterns on F-shape pages.
We will utilize these findings to design our F-shape click model (FSCM) in Section~\ref{sec:framework} for multi-block mobile pages.

\section{Problem Formulation}


Given a query $q$, the information retrieval system returns an F-shape page, which can be decomposed into interleavings of vertical and horizontal blocks: $\mathcal{B}=[B_1^v,B_2^h,B_3^v,B_4^h,...,B_{n-1}^h,B_n^v]$, with $v$ and $h$ denoting vertical and horizontal blocks. Each block $B_t^{b}$ ($b\in\{v,h\}$) contains $m_t$ items, denoted as $\mathcal{I}_t^{b}=[i_{t,j}^{b}]_{j=1}^{m_t}$. The user browses the page and clicks items that she is satisfied with. We define the click variable $c_{t,j}$ for each item $i_{t,j}^{b}$, where $c_{i,j}=1$ if $i_{t,j}^{b}$ is clicked by the user and 0 if not. Then we can define the problem of click model tasks on multi-block mobile pages as follows:

For the $j$-th item in the $t$-th block ($i_{t,j}^{b}$) on the F-shape page $(\mathcal{B},\{\mathcal{I}_t\}_{t=1}^n)$ with query $q$, given the user's previous interactions $\mathcal{C}=[c_{1,1},c_{1,2},...,c_{t,j-1}]$, we would like to predict whether the item $i_{t,j}^{b}$ will be clicked by the user (i.e., the click variable $c_{t,j}$).

Hereinafter, as $b$ can be determined by the parity of $t$, we omit the superscript $b$ of $i_{t,j}^b$ for brevity when there is no ambiguity.

\begin{figure*}[t]
    \centering
    \includegraphics[width=0.95\textwidth]{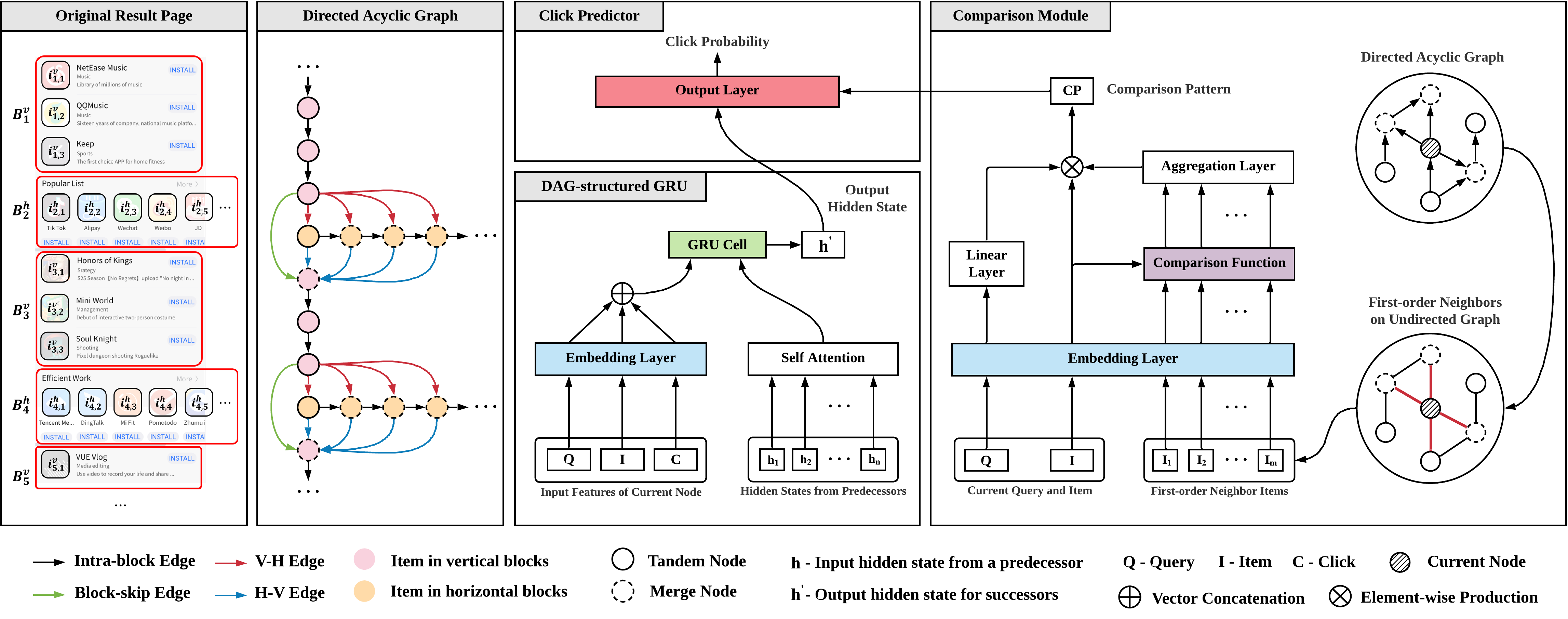}
    \vspace{-5pt}
    \caption{The framework of FSCM. We construct a directed acyclic graph (DAG) for an F-shape page, where each item is regarded as a vertex and each edge indicates the user's possible examination flow. The DAG-structured GRU and comparison module are proposed to model users' sequential (sequential browsing, block skip) and non-sequential (comparison) behaviors respectively.
    }
    \label{fig:framework}
\end{figure*}

\section{Model Framework}
\label{sec:framework}

As shown in Figure~\ref{fig:framework}, in this section, we introduce the framework of F-shape Click Model (FSCM), which serves as a general solution for click model tasks on multi-block mobile pages.

\subsection{Overview of FSCM}

We have conducted eye-tracking experiments to obtain users' \emph{sequential browsing}, \emph{block skip} and \emph{comparison} patterns towards F-shape pages in Section~\ref{sec:eye tracking}. These findings provide insights of user behaviors on multi-block mobile pages and lead to the design of FSCM. 
As shown in Figure~\ref{fig:framework}, we first construct a directed acyclic graph (DAG) for an F-shape page, where each item on the page is regarded as a vertex and each directed edge indicates the user's possible examination flow. DAG-structured gated recurrent units (GRUs) are applied on the nodes to model users' sequential browsing and block skip behaviors. Moreover, a comparison module is proposed to address users' comparison patterns, where the first-order neighbor items are considered as contexts when predicting the current item $i_{t,j}$. The click predictor combines the output hidden state and comparison pattern for each item, and predicts the click probability $p_{t,j}$.

Next, we will first introduce the graph construction method and general embedding layers, and then elaborate on the details of each module (i.e., DAG-structured GRU, comparison module, and click predictor). Moreover, we will discuss the extension of FSCM for general multi-block mobile pages.

\subsection{Graph Construction}

We aim to construct a directed acyclic graph (DAG) for each F-shape page, where each item is regarded as a vertex and each directed edge indicates the user's possible examination flow during the sequential browsing.
In Section~\ref{sec:eye tracking}, by answering \textbf{RQ1} (position bias), we have identified that users' general examination flows on an F-shape page are from top to bottom and from left to right. Furthermore, in \textbf{RQ2} (block skips), we find that V-V 2-length block skips are dominant among users' block-skip behaviors. Therefore, as shown in Figure~\ref{fig:framework}, the DAG consists of four kinds of directed edges:

\begin{itemize}[leftmargin=10pt]
    \item \textbf{Intra-block Edge} $(i_{t,j-1}^b,i_{t,j}^b),b=v \; \text{or} \; h, t=1,...,n,j=2,...,m_t$.
    The edges between each pair of two consecutive items in a block, which denotes the intra-block position bias and sequential examinations inside each block.
    \item \textbf{V-H Edge} $(i_{t,m_t}^v,i_{t+1,j}^h),t=1,3,5...,n,j=1,...,m_{t+1}$. 
    The edges from the last item in a vertical block to each item in the next horizontal block, which describes the block-level position bias and sequential examinations across blocks.
    \item \textbf{H-V Edge} $(i_{t,j}^h,i_{t+1,1}^v),t=2,4,6,...,n-1,j=1,...,m_t$. 
    The edges from each item in a horizontal block to the first item in the next vertical block, which denotes block-level position bias. Note that edges from the end of a horizontal block to the next vertical block may break the rule of "from left to right". But we also include them since "from top to bottom" is the primary principle.
    \item \textbf{Block-skip Edge} $(i_{t,m_t}^v,i_{t+2,1}^v),t=1,3,5,...,n-2$. 
    The edges from the last item in a vertical block to the first item in the next vertical block, which denotes the dominant V-V 2-length block-skip behaviors during the sequential browsing.
\end{itemize}

The graph construction method above generally implies users' examination flows when browsing the F-shape pages, which helps model users' sequential behaviors (DAG-structured GRU), as well as non-sequential behaviors (comparison module). 

\vspace{-3pt}
\subsection{Embedding Layer}

FSCM takes query $q$, item $i$ and (previous) click $c$ as inputs. Before the main process of the model, the original features are transformed into high-dimensional sparse features via one-hot encoding. Then we apply embedding layers on the one-hot vectors to map them to low-dimensional dense embedding vectors:
\begin{equation}
\begin{aligned}
    \mathbf{v}_{q}=\mathbf{Emb}_{\mathbf{q}}\left(q\right),
    \mathbf{v}_{i}=\mathbf{Emb}_{\mathbf{i}}\left(i\right),
    \mathbf{v}_{c}=\mathbf{Emb}_{\mathbf{c}}\left(c\right),
\end{aligned}
\end{equation}
where $\mathbf{Emb}_{\mathbf{*}} \in \mathcal{R}^{N_{*} \times l_{*}},\; * \in \{\mathbf{q},\mathbf{i},\mathbf{c}\}$. 
$N_*$ and $l_*$ denote the input feature size and embedding size. For ease of presentation, we omit the subscripts of embeddings when there is no ambiguity. If there are multiple feature fields for the query or item, we can obtain feature vectors for each field via similar embedding layers and concatenate them as the embedding of the query or item.

\vspace{-3pt}
\subsection{DAG-structured GRU}
\label{sec:DAG structured GRU}

After constructing the DAG that indicates users' examination flows, we aim to apply GRU operations to the graph, which is named DAG-structured GRU. A typical GRU cell~\cite{chung2014empirical} takes a feature vector $x_t$ and a previous hidden state $h_{t-1}$ as inputs, and generate a new hidden state $h_t=\operatorname{GRUcell}(x_t,h_{t-1})$.

We formulate the GRU cell operation as $h_{t,j}^{'}=\operatorname{GRUcell}(x_{t,j},h_{t,j})$ for each node $i_{t,j}$ on the DAG. The feature vector $x_{t,j}$ is the concatenation of embeddings of corresponding query, item and previous click. The input hidden state $h_{t,j}$ is aggregated from its predecessors via a self-attention layer. The output hidden state $h_{t,j}^{'}$ is then distributed to its successors to serve as their input hidden state. Hence, the DAG-structured GRU operation for each node $i_{t,j}$ is:
\begin{equation}
\begin{aligned}
    &x_{t,j}=[\mathbf{v}_{q} \oplus \mathbf{v}_{i_{t,j}} \oplus \mathbf{v}_{c_{t,j}}], \\
    &\alpha_{k} = \operatorname{Softmax}_k(\operatorname{MLP}(h_k^{'})), \;i_k \in \mathcal{P}_{t,j}, \\
    &h_{t,j}= \sum\nolimits_{i_k \in \mathcal{P}_{t,j}} \alpha_k h_k^{'}, \\
    &h_{t,j}^{'}=\operatorname{GRUcell}(x_{t,j},h_{t,j}),
\end{aligned}
\end{equation}
where $\mathcal{P}_{t,j}$ denotes the predecessor set of item $i_{t,j}$ and $\alpha_k$ is calculated through a shared two-layer Multi-Layer Perceptron (MLP). It is worth noting that the information of the current item $i_{t,j}$ is not utilized in the self-attention layer, since the distribution of the user’s previous examination is only affected by her actions on previous items - examination distribution is not affected by the content of the current item (users read the content only when the examination behavior happens). This temporal consideration has been verified by previous works~\cite{chen2020context,lin2021graph}.

Up to now, we have described the details of DAG-structured GRU from a unified view. 
Moreover, we argue that there are different underlying distributions of features and click signals for nodes (i.e., items) with different properties, thus requiring different GRU operation units with individual model parameters to extract the examination patterns. As shown in Figure~\ref{fig:framework}, we classify the nodes from two aspects as follows.

We classify the nodes into vertical nodes and horizontal nodes according to the blocks they belong to. In Section~\ref{sec:eye tracking}, we observe that the click distributions greatly differ between vertical and horizontal blocks due to several reasons (e.g., presentation bias, block skips), which need different GRU operation units respectively.

We classify the nodes into tandem nodes and merge nodes according to their graph indegrees. A node is defined as a merge node if its indegree is larger than 1. Otherwise, it is defined as a tandem node. Merge nodes serve as convergent points of users' distributed examination flows, which requires separate modelings. Also, since tandem nodes only have one predecessor, we can replace the self-attention layer with identity mapping to save memory.


Therefore, we maintain four different GRU cell operation units (each unit has individual model parameters) for (1) vertical tandem node, (2) vertical merge node, (3) horizontal tandem node, (4) horizontal merge node. Ablation studies in Section~\ref{sec:ablation} validate the effectiveness of these two node classification perspectives. By performing DAG-structured GRU, we can obtain output hidden state $h_{t,j}^{'}$ for each item on an F-shape page, indicating the user's sequential browsing presentation.

\subsection{Comparison Module}

We model users' sequential browsing and block skips by applying DAG-structured GRUs. Our eye-tracking experiments and previous work~\cite{zhang2021constructing} verify the necessity of considering non-sequential behaviors, especially comparison patterns. Hence, we propose the comparison module to model users' comparison behaviors.

Comparison behaviors can be considered as quick shifts of attention between two items. Therefore, every single predecessor or successor of the current node $i_{t,j}$ can be regarded as its comparison candidate. As shown in Figure~\ref{fig:framework}, we first convert the DAG into an undirected version by removing the direction of each edge, and collect first-order neighbors of the item to form its comparison candidate set $\mathcal{N}_{t,j}$. Then we perform a comparison function between the current item $i_{t,j}$ and candidate items $i_k$ in the comparison candidate set $\mathcal{N}_{t,j}$, which is followed by an aggregation layer to generate the comparison context presentation $cp_{t,j}$:
\begin{equation}
\begin{aligned}
    &\gamma_k =  \operatorname{Softmax}_k (g(\mathbf{v}_{i_{t,j}},\mathbf{v}_{i_{k}})), \; i_k \in \mathcal{N}_{t,j}, \\
    & cp_{t,j} = \sum\nolimits_{i_k \in \mathcal{N}_{t,j}} \gamma_k \mathbf{v}_{i_{k}},
\end{aligned}
\end{equation}
where $g(\cdot,\cdot)$ is the comparison function. We implement three different comparison functions:
\begin{enumerate}
    \item \textbf{Inner product}: $g(v_1,v_2)=v_1^Tv_2$.
    \item \textbf{Neural function}: $g(v_1,v_2)=MLP([v_1 \oplus v_2])$.
    \item \textbf{Kernel function}: $g(v_1,v_2)=v_1^T W v_2$.
\end{enumerate}

After comparison and aggregation, we further apply element-wise product on the comparison context presentation $cp_{t,j}$, current item embedding $\mathbf{v}_{i_{t,j}}$ and query embedding $\mathbf{v}_{q}$ (a linear layer is performed to ensure dimensionality consistency):
\begin{equation}
\begin{aligned}
    cp_{t,j}^{'}=cp_{t,j} \odot \mathbf{v}_{i_{t,j}} \odot \mathbf{v}_{q}
\end{aligned}
\end{equation}
where $\odot$ is element-wise product. Now we obtain the comparison pattern $cp_{t,j}^{'}$ for each item $i_{t,j}$, which indicates the user's primary non-sequential behaviors.

\subsection{Click Predictor}

The click predictor combines GRU hidden state $h_{t,j}^{'}$ and comparison pattern $cp_{t,j}^{'}$, and outputs the click probability. As shown in Figure~\ref{fig:framework}, for item $i_{t,j}$, we concatenate $h_{t,j}^{'}$ and $cp_{t,j}^{'}$ together, and put them through an output layer to predict the click probability $p_{t,j}$:
\begin{equation}
    p_{t,j}=\operatorname{Sigmoid}(\operatorname{MLP}([h_{t,j}^{'} \oplus cp_{t,j}^{'}])).
\end{equation}

After click prediction, we adopt a binary cross-entropy loss to ensure an end-to-end training of FSCM. The objective function to be minimized during training is:
\begin{equation}
    \mathcal{L}(\theta)=\mathcal{L}_{c}(\theta)+\lambda \| \theta \|^2
\end{equation}
\begin{equation}
    \mathcal{L}_c(\theta)=-\frac{1}{N}\sum_{k=1}^{N}\sum_{t,j}C_{t,j}^k\log \mathcal{P}_{t,j}^k+(1-C_{t,j}^k)\log(1-\mathcal{P}_{t,j}^k)
\end{equation}
where $\theta$ denotes trainable parameters in FSCM, $\lambda$ denotes the hyperparameter for regularization, $N$ denotes the number of training batches, $\mathcal{C}_{t,j}^k$ and $\mathcal{P}_{t,j}^k$ denote the real click signal and the predicted click probability in the $k$-th training batch.

\subsection{Extension of FSCM on General Multi-block Mobile Pages}

\begin{figure}[t]
    \centering
    \includegraphics[width=0.47\textwidth]{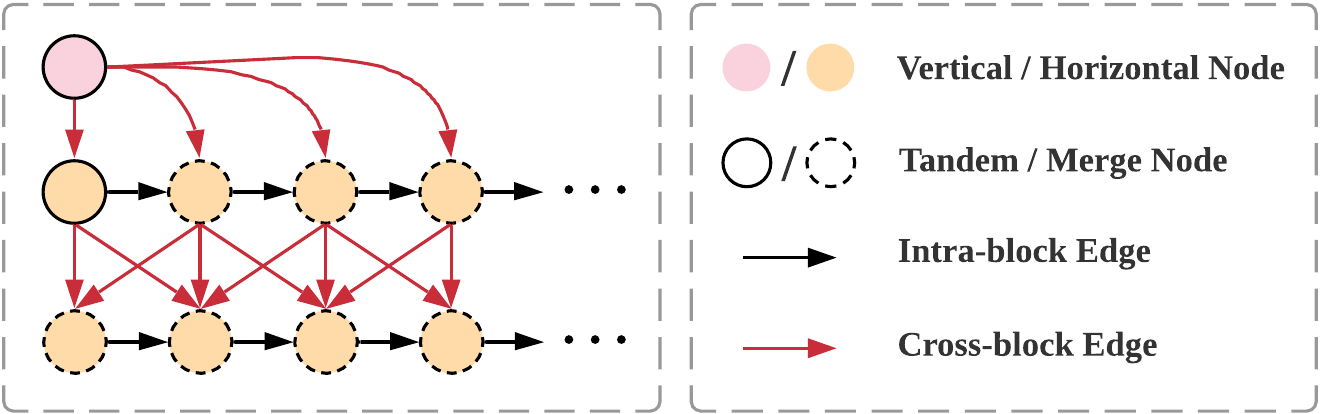}
    \caption{An example of DAG construction for FSCM extension on general multi-block mobile pages.}
    \label{fig:extension}
\end{figure}

The proposed FSCM above focuses on the F-shape pages, which can be decomposed into interleavings of vertical and horizontal blocks. Although F-shape display forms are prevalent for multi-block mobile pages, there exist presentation styles that break the F-shape rule (e.g., multiple consecutive horizontal blocks).

We argue that FSCM can be flexibly extended to any multi-block mobile pages by making some preliminary assumptions on users' browsing behaviors. The assumptions may be concluded from a careful eye-tracking study, or come from simple statistics and data analysis. Then, following the rationale presented above, we can construct a DAG according to the assumptions, which indicates users' possible examination flows on that kind of multi-block mobile page. An example of DAG construction for two consecutive horizontal blocks is shown in Figure~\ref{fig:extension}. Next, DAG-structured GRU and comparison module can be directly applied to model users' behaviors across blocks, showing the flexibility and the generalization ability of FSCM. 


\section{Experiment}
\label{sec:exp}

In this section, we conduct extensive experiments to answer the following questions:

\begin{itemize}
    \item[\textbf{RQ1}] Does FSCM achieve the best performance compared to the baseline models?
    \item[\textbf{RQ2}] How do different comparison functions in the comparison module influence the prediction performance?
    \item[\textbf{RQ3}] What is the influence of different components in FSCM?
\end{itemize}

\subsection{Experimental Setup}

\subsubsection{Dataset}

Since there are no public datasets for multi-block mobile pages, we collect \textbf{AppStore} dataset from a mainstream commercial App Store with F-shape pages, from September 17, 2021 to November 14, 2021. 
The first $54$ days are training set, while the last $5$ days are randomly split into valid set and test set. 
The user behavioral history is collected in real-time.
We discard queries that have no positive interactions (clicks) with the retrieval system.
The dataset consists of $394046$ unique queries and $1646$ items (Apps). 
Each item contains $7$ fields of features and each query contains $26$ fields of features.
The dataset statistics are shown in Table \ref{tab:dataset}. 

\begin{table}[thb]
    \centering
    \vspace{-3pt}
    \caption{The dataset statistics. A session corresponds to an F-shape page. A block indicates a vertical or horizontal list.}
    \vspace{-3pt}
	\resizebox{0.45\textwidth}{!}{
    \begin{tabular}{cccc}
     \toprule
     & training & validating & testing\\
     \midrule
    $\#$ sessions & $637959$ & $40792$ & $40792$ \\
    avg. blocks per session & 3.6654 & 3.7392 & 3.7374 \\
    avg. clicks per session & 0.6510 & 0.7248 & 0.7313 \\
    \bottomrule
    \end{tabular}
    }
    \vspace{-5pt}
    \label{tab:dataset}
\end{table}

\subsubsection{Baselines}
The mainstream click models could be classified into probabilistic graphical model (PGM) based and neural network (NN) based methods. We do not adopt PGM-based methods as baselines due to the following two reasons:

\begin{itemize}[leftmargin=10pt]
    \item The AppStore dataset contains multiple feature fields for queries and items, but classical PGM-based methods can only deal with query ID and item ID inputs, which definitely lead to inferior performance.
    \item PGM-based methods require every query ID or item ID in the test set to appear in the training set. Otherwise they fail to make reasonable inferences. But AppStore dataset is too sparse to meet this requirement (1.0638 sessions per query ID on average).
\end{itemize}

Moreover, existing multi-block click models for desktop contexts (e.g., JRE~\cite{srikant2010user}, GCM~\cite{zhu2010novel}, WPC~\cite{chen2011whole}) are all based on PGM framework. Hence, we only compare our model with single-list NN-based click models. We choose NCM~\cite{borisov2016neural} and GraphCM~\cite{lin2021graph} as representative baselines, and extend them into multi-block versions in two different ways:

\begin{itemize}[leftmargin=10pt]
    \item \textbf{Block-wise}. We maintain two independent click models for vertical and horizontal blocks respectively. Two separate models are updated simultaneously towards the same session data.
    \item \textbf{List-wise}. We also maintain two independent click models for vertical and horizontal blocks respectively, and update them simultaneously on the same session data. The difference is that the list-wise version concatenates all the vertical blocks and views them as a single vertical list on the page, which is similar to block-skip edges of the DAG constructed in FSCM.
\end{itemize}


\subsubsection{Evaluation Metrics}
Following previous works~\cite{chuklin2015click}, we use the log-likelihood (LL) and the area under the ROC curve (AUC) as evaluation metrics, which reflect point-wise likelihood and pairwise ranking performance respectively. Higher values of LL and AUC correspond to better click prediction performance of a click model. 

It is worth noting that we do not adopt perplexity (PPL) as the evaluation metric. For AppStore dataset, we notice that the length of each block varies, and the number of items displayed in the top-ranked positions is larger than that in the low-ranked positions. PPL metric requires to first average the perplexity for items at the same position, and then average perplexity over different positions. Therefore, in AppStore dataset, PPL would emphasize predictions at low-ranked positions since the numbers of items at lower ranks are smaller, which is irrational and violates the position bias.

\begin{table}[t]
	\centering
	\caption{Overall performance of click models. The best results are given in bold, while the second best values are underlined. B.W/L.W. mean block-wise/list-wise implementation. Rel. Impr. means relative improvement of FSCM against the best baseline. Improvements are all statistically significant with $p< 10^{-6}$, where LL is measured by t-test and AUC is measured on an online website\href{http://vassarstats.net/roc_comp.html}{[\emph{link}]} under the guidance of~\cite{hanley1982meaning}.}
	\label{tab:experiment}
	\resizebox{0.48\textwidth}{!}{
	\renewcommand\arraystretch{1.3}
	\begin{tabular}{c| c|c c c|c c c}
	    \toprule
		\hline
		\multicolumn{2}{c|}{\multirow{2}{*}{Model}} &  \multicolumn{3}{c|}{\textbf{LL}}  &  \multicolumn{3}{c}{\textbf{AUC}} \\
		\cline{3-8}
		\multicolumn{2}{c|}{} & Vertical & Horizontal & Overall  & Vertical & Horizontal & Overall\\
		\cline{1-8}
		\multirow{2}{*}{NCM} & 
		\multicolumn{1}{c|}{B.W.} & $-0.1087$ &$-0.1790$& $-0.1504$& $0.5127$& $0.6485$& $0.6326$\\
		& \multicolumn{1}{c|}{L.W.} & $-0.1084$ & $-0.1608$& $-0.1421$&  $0.6389$& $0.7004$& $0.6899$\\
		\cline{1-2}
		\multirow{2}{*}{GraphCM} &\multicolumn{1}{c|}{B.W.} & $-0.1625$& $-0.1643$& $0.9625$& $0.5589$& $0.6679$& $0.6373$\\
		&\multicolumn{1}{c|}{L.W.} & $\underline{-0.1019}$ & $\underline{-0.1574}$ & $\underline{-0.1348}$  & $\underline{0.6927}$& $\underline{0.8092}$& $\underline{0.7884}$\\
		\cline{1-8}
		\multicolumn{2}{c|}{FSCM} & \textbf{-0.0757} & \textbf{-0.1360} & \textbf{-0.1113} & \textbf{0.7519}& \textbf{0.8456}& \textbf{0.8350}\\
		\cline{1-8}
		\multicolumn{2}{c|}{Rel. Impr.} & $25.71\%$& $13.59\%$& $17.43\%$& $8.55\%$ &$4.50\%$ & $5.91\%$\\
 		\hline
 		\bottomrule
	\end{tabular}
	}
\end{table}

\subsubsection{Implementation Details}
We train FSCM with a batch size of 1024 using Adam optimizer. 
The embedding size of each feature field is $4$.
The hidden sizes of GRU cells are $128$.
The learning rate is $0.001$.
To ensure fair comparison, we also fine-tune all the baseline models to achieve their best performance.
The implementation of our model is available\footnote{The TensorFlow implementation is available at: \url{https://github.com/fulingyue/F-shape-Click-Model-FSCM.git}. The MindSpore implementation is available at: \url{https://gitee.com/mindspore/models/tree/master/research/recommend/FSCM}}.

\subsection{Performance Comparison (RQ1)} 

We perform click prediction task on each click model for performance comparison. The results are shown in Table~\ref{tab:experiment}, from which we could obtain the following observations:

\begin{itemize}[leftmargin=18pt]
    \item[(1)] List-wise models achieve better performance compared to block-wise models in both metrics. List-wise methods link all the vertical blocks and treat them as a single vertical list, which is similar to block-skip edges of the DAG in FSCM (V-V 2-length block skips). This validates the effectiveness to model user behaviors across blocks instead of in an isolated manner.
    \item[(2)] List-wise GraphCM achieves the best performance among the baseline models. Compared to NCM, GraphCM considers the relationship between examination, relevance and clicks, where the examination hypothesis is encoded explicitly. GraphCM also adds the interaction of adjacent items, which helps the model to capture local structural information.
    \item[(3)] FSCM outperforms all the baseline models on both metrics by a large margin. FSCM achieves page-level click prediction by extracting useful dependencies among user examination flows via DAG-structured GRUs. A comparison module is proposed to further model users' comparison patterns. These factors lead to significant improvement over baselines. Such improvement also validates that the underlying assumptions and model structures of FSCM are closer to practical user behaviors than the assumptions made in competing models.
    \item[(4)] The improvement of FSCM is more significant in vertical blocks than horizontal blocks. According to our eye-tracking experiments, users tend to pay more attention to the vertical blocks due to presentation bias and possibly give more click feedback, which makes it easier to capture user behavior patterns towards vertical blocks. Moreover, block-skip edges of the DAG in FSCM provide explicit modelings of users' V-V 2-length block-skip behaviors, thus better promoting the performance on vertical blocks.
\end{itemize}

\subsection{Comparison Function (RQ2)}

\begin{figure}[tb]
     \centering
     \includegraphics[width=0.4\textwidth]{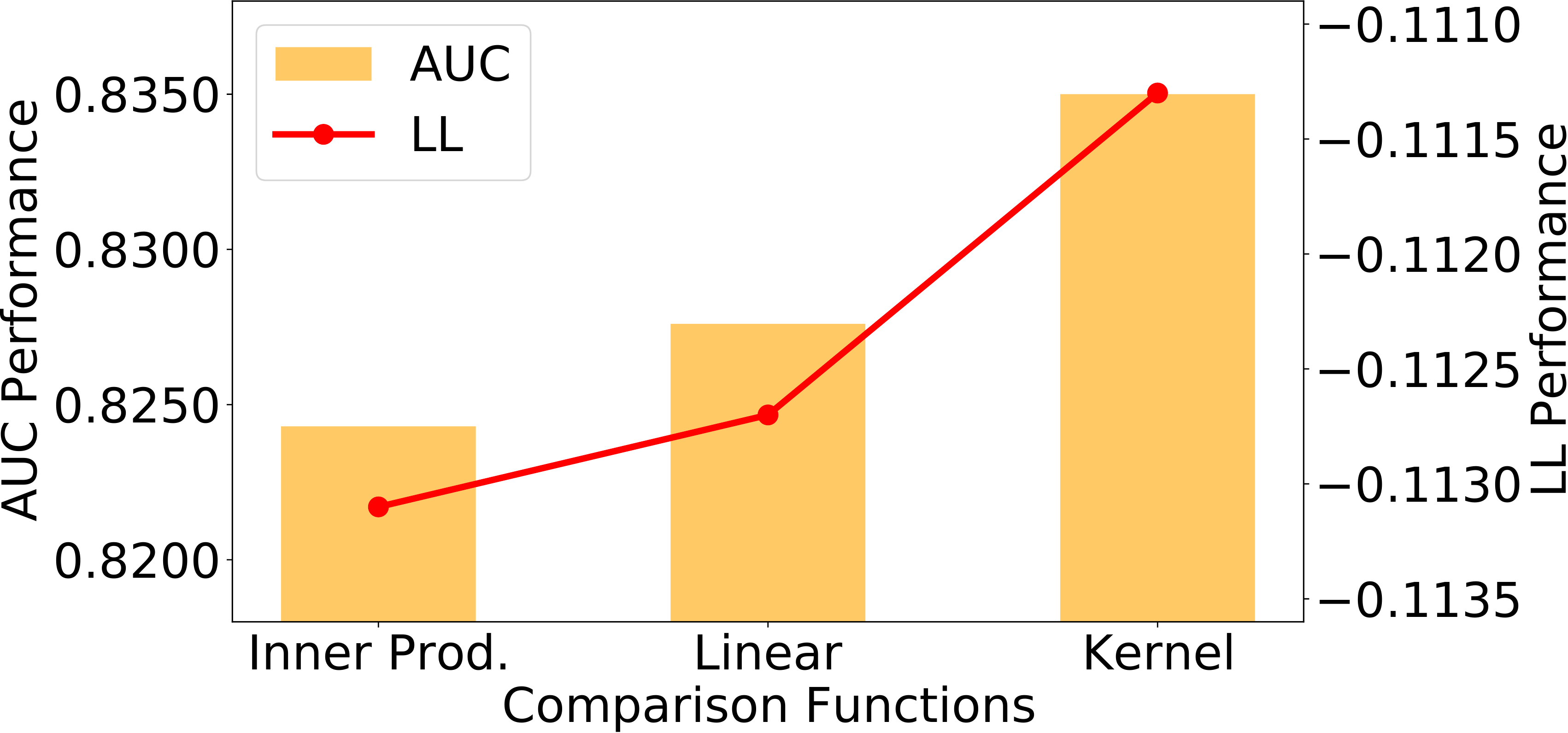}
     \vspace{-7pt}
     \caption{FSCM's click prediction performance with different comparison functions. (Best viewed in color.)}
     \vspace{-10pt}
     \label{fig:comparison function}
\end{figure}

We study the effectiveness of different comparison functions by comparing their click prediction performance, which is shown in Figure~\ref{fig:comparison function}. We observe that neural and kernel comparison functions outperform the inner product function, which validates that introducing additional learnable parameters could better capture users' comparison patterns between items. Besides, the kernel function achieves better performance than the neural function which is implemented as a two-layer fully connected network. We argue that the kernel function applies explicit second-order feature interactions between items, which augments the feature engineering and is more 
consistent with users' comparison patterns, thus resulting in better performance.

\subsection{Ablation Study (RQ3)}
\label{sec:ablation}

\begin{table}[t]
	\centering
	\caption{Ablation study on FSCM. We perform the following operations respectively: (\romannumeral1) remove comparison module; (\romannumeral2) remove block-skip edges when constructing DAGs;
	(\romannumeral3) share model parameters between vertical and horizontal nodes for DAG-structured GRUs;
	(\romannumeral4) share model parameters between tandem and merge nodes for DAG-structured GRUs;}
	\label{tab:ablation}
	\resizebox{0.49\textwidth}{!}{
	\renewcommand\arraystretch{1.3}
	\begin{tabular}{c|c c c|c c c}
	    \toprule
		\hline
		\multicolumn{1}{c|}{\multirow{2}{*}{Model}} &  \multicolumn{3}{c|}{\textbf{LL}} &  \multicolumn{3}{c}{\textbf{AUC}} \\
		\cline{2-7}
		\multicolumn{1}{c|}{} & Vertical & Horizontal & Overall  & Vertical & Horizontal & Overall\\
		\cline{1-7}
		\multicolumn{1}{c|}{0. FSCM} & \textbf{-0.0757} & \textbf{-0.1360} & \textbf{-0.1113} & \textbf{0.7519}& \textbf{0.8456}& \textbf{0.8350}\\
		\multicolumn{1}{c|}{\romannumeral1. w/o Comp.} & $-0.0773$ & $-0.1369$ & $-0.1126$ & $0.7197$ & $0.8437$& $0.8283$\\
		\multicolumn{1}{c|}{\romannumeral2. w/o Skip.} & $-0.0768$ & $-0.1370$ & $-0.1124$ &  $0.7278$& $0.8428$& $0.8289$\\
 		\multicolumn{1}{c|}{\romannumeral3
 		. share H/V} &$-0.0790$& $-0.1376$& $-0.1137$& $0.7013$& $0.8412$& $0.8232$\\
 		\multicolumn{1}{c|}{\romannumeral4. share T/M} &$-0.0775$& $-0.1376$& $-0.1130$& $0.7224$& $0.8415$& $0.8267 $\\
 		\hline
 		\bottomrule
	\end{tabular}
	}
	\vspace{-10pt}
\end{table}
In order to investigate the contribution of different components to the final performance of FSCM, we conduct several comparison experiments by performing the following operations: (\romannumeral1) remove the comparison module; (\romannumeral2) remove the block-skip edges when constructing DAGs; (\romannumeral3) share the model parameters between vertical and horizontal nodes for DAG-structured GRUs; (\romannumeral4) share the model parameters between tandem and merge nodes for DAG-structured GRUs. The results are reported in Table~\ref{tab:ablation}.


When we remove the comparison module, the performance on both metrics degrades, which suggests that it is crucial to model non-sequential behaviors when designing a click model, and our comparison module can extract users' comparison patterns to help user behavior predictions.
Likewise, the performance drops when we remove block-skip edges of the DAG. 
Block-skip edges provide direct modelings of users' V-V 2-length block skips, which are observed as the primary skip behaviors in our eye-tracking experiments. 
These facts validate that we can enhance the performance of click models by making proper preliminary assumptions. For instance, in FSCM, we make explicit modelings of block skips and comparison patterns in addition to users' basic sequential browsing.

In FSCM, after constructing the DAG, we classify the nodes from two aspects (i.e., vertical/horizontal, and tandem/merge) and maintain four different DAG-structured GRU operation units with individual model parameters. As shown in Table~\ref{tab:ablation}, the performance worsens when we discard one of the node classifications and share the GRU parameters between vertical and horizontal nodes (or tandem and merge nodes). This demonstrates the effectiveness of our node classification perspectives, and there are different underlying distributions of features and click signals for nodes (i.e., items) with different properties, which requires individual GRU operation units to achieve better performance.


\section{Conclusion}

In this paper, we conduct a lab-based eye-tracking study to thoroughly investigate user behaviors on F-shape pages and obtain users' sequential browsing, block skip and comparison patterns.
These findings lead to the design of a novel F-shape Click Model (FSCM), which serves as a general solution to user modelings on multi-block mobile pages.
We first construct a DAG for a result page, where each item is regarded as a vertex and each edge indicates the user's possible examination flow.
Next, DAG-structured GRUs and a comparison module are applied to explicitly model users' behaviors and interactive influence across different blocks.
The experimental results on a large-scale real-world dataset validate the effectiveness of FSCM on user behavior predictions compared with baseline click models.
For the future work of research, we plan to utilize FSCM in the downstream tasks (e.g., offline evaluation and page-level ranking policy).
\section*{Acknowledgement}

The SJTU team is supported by Shanghai Municipal Science and Technology Major Project (2021SHZDZX0102) and National Natural Science Foundation of China (62177033). The work is also sponsored by Huawei Innovation Research Program.
We thank MindSpore~\cite{mindspore} for the partial support of this work.

\bibliographystyle{ACM-Reference-Format}
\bibliography{acmart}

\end{document}